\documentclass[runningheads]{llncs}
\usepackage[T1]{fontenc}
\usepackage{booktabs}
\usepackage{makecell}
\usepackage[numbers,sort]{natbib}
\usepackage{graphicx}
\begin{document}
\title{Contribution Title}
%
%

\title{Building a Media Ecosystem Observatory from Scratch: Infrastructure, Methodology, and Insights}
\title{Building a Media Ecosystem Observatory from Scratch: A Data Infrastructure for Cross-Platform Analysis}
\author {
    Zeynep Pehlivan \and
    Saewon Park  \and
    Alexei Sisulu Abrahams  \and
    Mika Desblancs-Patel  \and
    Benjamin David Steel  \and
    Aengus Bridgman
}
\authorrunning{Z.Pehlivan et al.}
%

\institute{McGill University, Montreal, Canada, 
\email{\{zeynep.pehlivan,saewon.park,alexei.abrahams,mika.desblancs,\\
benjamin.steel,aengus.bridgman\}@mcgill.ca}}

\maketitle              
\begin{abstract}
Understanding the flow of information across today’s fragmented digital media landscape requires scalable, cross-platform infrastructure. In this paper, we present the Canadian Media Ecosystem Observatory, a national-scale infrastructure designed to monitor political and media discourse across platforms in near real time.

Media Ecosystem Observatory(MEO) data infrastructure features custom crawlers for major platforms, a unified indexing pipeline, and a normalization layer that harmonizes heterogeneous schemas into a common data model. Semantic embeddings are computed for each post to enable similarity search and vector-based analyses such as topic modeling and clustering. Processed and raw data are made accessible through API, dashboards and web site, supporting both automated and ad hoc research workflows. We illustrate the utility of the observatory through example analyses of major Canadian political events, including Meta’s 2023 news ban and the recent federal elections. As a whole, the system offers a model for digital trace infrastructure and an evolving research platform for studying the dynamics of modern media ecosystems.

\end{abstract}

\section{Introduction}
To understand how information, stories, and influence move online today, we need to look at more than just one platform at a time \cite{bossetta_cross-platform_2023,yarchi_political_2021}. Instead, we must think of the whole "information ecosystem", a network of connected communities across social media and news websites. These platforms have different demographic breakdowns, usage patterns, and affordances, but are highly interdependent; users and content frequently move from one to another. This broader view is useful, but building a system to study it is difficult. It requires strong technical tools, clever ways to work with different types of data, and teams from multiple disciplines.

One way to deal with this challenge is to build a digital media observatory. These include creating systems to collect data at scale, managing different data types in one place, giving access, developing tools for analysis and dealing with changes in how platforms allow access to data—such as the shutdown of CrowdTangle or changes to X/Twitter's API.

In Canada, while researchers have studied parts of the digital media environment, for further studies we need a clear picture of how political messages
or media stories move across platforms or how they affect public debate overall. Most studies still focus on only one platform at a time, which gives us only a part of the story \cite{muric_covid-19_2021,dogdu_detecting_2024, yang_regional_2024,Ai_Gupta_Oak_Hui_Liu_Hirschberg_2024, sun_social_2024}.

This paper introduces the Canadian Media Ecosystem Observatory. We explain how we built it from scratch by detailing: how we collected data from major social media platforms, how we processed and connected this data, and how we combined quantitative and qualitative methods to study online activity in Canada.

Our main contributions are:
\begin{itemize}
    \item A model for building a national media observatory in the face of limited access to platform data.
    \item A mixed-method approach for analyzing cross-platform data.
    \item Use cases showing how political conversations take place across different platforms in Canada.
\end{itemize}

\section{Related Work}
Building infrastructure for extensive, cross-platform social media research has gained popularity in recent years. Institutions, universities, and other organizations have launched a number of initiatives \cite{pehlivan2021archiving,Yang2022Botometer101,wiedemann_concept_2023,matassi_agenda_2021,ronzhyn_defining_2023}  to make it easier to gather, curate, and analyze digital trace data in a methodical manner for the study of media dynamics and political communication.

Several projects focus on the long-term monitoring of selected speaker groups on individual platforms. An in-depth discussion of the challenges of archiving social media content, focusing on the case of X/Twitter, is discussed in \cite{pehlivan2021archiving}. Their work highlights technical and ethical issues in long-term data preservation, particularly regarding the volatility and ephemerality of platform content. However, it does not address multi-platform integration or support advanced analytical pipelines such as semantic indexing.

Other large-scale efforts, such as the Observatory on Social Media (OSoMe) at Indiana University, have developed specialized tools for Twitter data, including bot detection (Botometer \cite{Yang2022Botometer101}) and misinformation tracking (Hoaxy \cite{Shao_2016}). These infrastructures have supported important research on information diffusion, especially during electoral cycles and public health crises. However, they remain largely focused on single-platform workflows.

In contrast, recent research has called for a comparative and cross-platform turn in media and communication studies. Wiedemann et al.~\cite{wiedemann_concept_2023} describe the development of the Social Media Observatory (SMO) as a DIY-oriented infrastructure for long-term monitoring of public communication on social platforms and online news media in Germany. Their open-science approach emphasizes curated speaker lists, reproducibility, and ethical considerations, offering tools and workflows to support decentralized data collection and analysis efforts. While our infrastructure shares many of these core principles, it places greater emphasis on near real-time collection, semantic indexing, and centralized programmatic access to enable automated, large-scale, and cross-platform analyses.

Other projects have contributed to the comparative turn in social media studies, encouraging researchers to consider the interplay of national context, platform affordances, and media systems \cite{matassi_agenda_2021, ronzhyn_defining_2023}. Hase et al. \cite{hase_adapting_2023} illustrate this by examining how German news organizations adapt their messaging across platforms like Facebook, TikTok, Instagram, and Twitter, revealing partial alignment with platform logics. Such work highlights the value of multi-platform analysis and motivates the need for infrastructures that enable longitudinal and cross-contextual studies.

In terms of infrastructure design, various studies have emphasized the value of combining raw data archives with modular, extensible processing pipelines. For instance,  an AI-enhanced media monitoring system for marketing applications, while emphasizing automated insight extraction and competitor tracking, is introduced in \cite{perakakis_social_2019}. Although focused on a different domain, the modular design principles are widely applicable and parallel many components of our system, including crawler orchestration, containerized processing, and dashboard-based access.

The Canadian Media Ecosystem Observatory described in this paper integrates digital trace data across multiple platforms, combines structured and unstructured content, and provides uniform, real-time access through dashboards, API and frontend solution. We see this work as a replicable model that balances standardized tooling with localized customization. Adapting the system for a different country primarily involves generating a new seed list of relevant entities (e.g., politicians, media outlets, influencers) and customizing metadata schemas where necessary. Our goal is to contribute to this growing ecosystem of media observatories by sharing the architectural, methodological, and practical lessons learned from building a national-scale, multi-platform infrastructure.

\section{System Overview}

The Canadian Media Ecosystem Observatory is designed as a modular and scalable infrastructure for collecting, storing, analyzing, and visualizing digital trace data from multiple social media platforms. The system supports both almost real-time (d-3) and historical data processing and is intended to provide flexible, cross-platform insights into Canadian information ecosystem.

\begin{figure}
    \centering
    \includegraphics[width=0.9\linewidth]{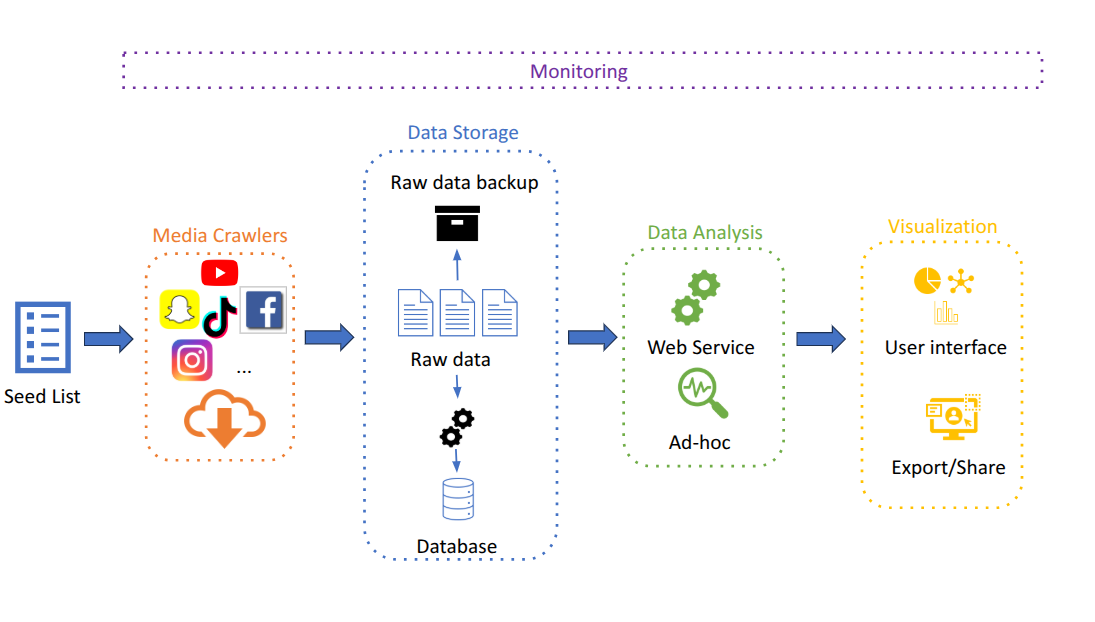}
    \caption{Media Ecosystem Observatory Overall Architecture}
    \label{fig:architecture}
\end{figure}

Figure \ref{fig:architecture} illustrates the overall architecture of the observatory. The observatory consists of five main components:

\begin{itemize}
    \item Seed List: defines and controls what data to collect;
    \item Media Crawlers: gathers data from different platforms
    \item Data Indexing: handles raw data ingestion, transformation, and backup;
    \item Data Access: supports automated and ad-hoc analysis pipelines and data access, provides access to the processed information through dashboards and export tools;
    \item Monitoring Layer: tracks system performance and data and pipeline health.
\end{itemize}

Each component is designed to be extensible and interoperable to ensure long-term sustainability under changing data policies and research needs. In the following subsections, we describe each component in detail.

\subsection{Seed List Creation and Annotations}
Our dataset consists of posts collected on X/Twitter, Instagram, TikTok, Facebook, Bluesky and YouTube from relevant actors in the Canadian media ecosystem. We refer to these actors as seeds, which we define as a person, group, organization, or media product that is of substantive interest to Canada's media ecosystem. Examples of media products are newspapers and podcasts. To be included in the seed list, these entities must be alive, in the case of a person, or in operation at the time of inclusion, for the remaining types. In addition, to remain included in the seed list, the entity must have activity on at least one platform in the last 12 months. 

Our goal was to generate as comprehensive a list as possible within the scope of national and provincial political actors, major media organizations, and relevant influencers. While we refer to this as “exhaustive” in scope, the seed list intentionally prioritizes national and provincial-level political actors, excluding, for example, municipal politicians from cities under 100,000 residents. We aim for high coverage of the most visible and impactful actors in Canadian political and media discourse, prioritizing signal over comprehensiveness at the long tail.

To identify politicians, we used official rosters of elected members of the House of Commons and all provincial legislative assemblies, collecting each politician’s name, province, and political party. Provincial parties were mapped to their federal counterparts to facilitate cross-level comparisons. Additional variables, such as electoral district information, were collected for federal and British Columbia provincial politicians in preparation for analyses of the 2025 federal and 2024 British Columbia provincial elections.

\begin{table}[htbp]
\centering

\label{tab:platform_entity_distribution}
\caption{Distribution of Entities Across Platforms}
\resizebox{\columnwidth}{!}{%
\begin{tabular}{@{}lrrrrrrr@{}}
\toprule
\textbf{Platform} & \textbf{\# Politician (\%)} & \textbf{\# News (\%)} & \textbf{\# Influencer (\%)} & \textbf{\# Gov. (\%)} & \textbf{\# CSO (\%)} & \textbf{Foreign}  & \textbf{Total} \\ 
\midrule
X/Twitter         & 1,459 (73.1\%)  & 699 (81.0\%)   & 1,031 (87.4\%)  & 192 (93.2\%)  & 653 (87.2\%)  & 177 (94.1\%)  & 4,211 \\
Instagram         & 1,450 (72.6\%)  & 383 (44.4\%)   & 597 (50.6\%)    & 113 (54.9\%)  & 624 (83.3\%)  & 87 (46.3\%)   & 3,254 \\
YouTube           & 295 (14.8\%)    & 142 (16.5\%)   & 186 (15.8\%)    & 131(63.6\%)   & 510 (68.1\%)  & 68 (36.2\%)   & 1,332   \\
TikTok            & 125 (6.3\%)     & 63 (7.3\%)     & 214 (18.2\%)    & 2 (1.0\%)     & 134 (17.9\%)  & 16 (8.5\%)    & 554   \\ 
Facebook          & 1,728 (86.5\%)  & 793 (91.9\%)   & 397 (33.7\%)    & 156 (75.7\%)  & 700 (93.5\%)  & 102 (54.3\%)  & 3,876   \\ 
Telegram          & 1 (0.1\%)       & 1 (0.1\%)      & 27 (2.3\%)      & --            & 5 (0.7\%)     & 55 (29.3\%)   & 89   \\
Bluesky           & 217 (10.9\%)    & --             & 188 (15.9\%)    & --            & --            & --            & 405    \\ 
\midrule
\textbf{Total}    & 5,275           & 2,081          & 2,640           & 594           & 2,626         & 505           & 13,721 \\
\bottomrule
\end{tabular}
}

\label{tab:platform_entity_distribution}
\end{table}

For the identification of news outlets, we used Media Cloud’s\footnote{https://www.mediacloud.org/} “national” and “state-local” datasets of Canadian news outlets. Media Cloud’s lists were manually reviewed to confirm that they were still in operation and based in Canada. During this process, we also determined if the outlet was “national”, meaning that the outlet covers the majority of Canada’s provinces and territories, or “local”, meaning that the outlet covers a minority of Canada’s provinces and territories. This categorization was recorded as a sub-type of the entity. If a news outlet was labeled as “local”, its province information was added to the seed list based on the location of its headquarters.

We attempted to identify as many political influencers who met the following criteria: at least 10,000 followers in TikTok/X or at least 5,000 followers in Facebook/Instagram/Youtube/Bluesky, at least a 1:2 following to followers threshold, and have a majority of the content be political. For government organizations, we consulted the official website of the Government of Canada for the list of federal departments and agencies. For civil society organizations, we consulted Revenue Canada's charities list and the Government of Canada's list of unions and filtered only organizations that met the follower threshold or revenue and membership thresholds. Lastly, we identified major state actors, media, and influencers from a select number of foreign countries (USA, Russia, India, and China) to study potential foreign influence on Canada’s media ecosystem\footnote{\url{https://foreigninterferencecommission.ca/fileadmin/report\_volume\_1.pdf}}.

Each entity was annotated with attributes such as main type, subtype, party affiliation (provincial and federal), electoral riding, and province. Verified social media handles across all available platforms were collected, and collection tags were assigned to aid in filtering. Only official accounts posting political or news-related content were included; personal or non-relevant accounts were excluded. In total, our seed list comprises 5,515 unique entities. Table~\ref{tab:platform_entity_distribution} shows the distribution of these entities by platform and type, including the proportion of politicians and news outlets with accounts on each platform.



\subsection{Media Crawlers}
To build a comprehensive dataset, we developed platform-specific tools and workflows for collecting data from social media platforms. Metadata for each post was preserved in its entirety and stored in JSON format to ensure structural consistency and completeness. While we did not collect media files directly, all available links to media content (e.g., images, videos) were retained in the metadata. This section outlines the specific data collection strategies implemented for each platform:

\subsubsection{X/Twitter}
We developed a web scraping library for X/Twitter that programmatically queries the platform’s advanced search and timeline endpoints to retrieve recent posts. This approach allowed us to collect data on a bi-weekly basis, as well as historical posts from the months prior to an account’s inclusion in our seed list, by specifying custom date ranges. Data was collected directly from structured API-like responses rather than rendered web pages, ensuring completeness and consistency of post metadata across the dataset.

\subsubsection{TikTok}
We used a web scraping library for TikTok as described in \cite{steel2024invasion} that allows us to do full reverse-chronological collections for each account. The scraper was run at regular weekly intervals to ensure comprehensive coverage of posts while maintaining the integrity of engagement metrics. While an official TikTok Research API does exist, it does not provide data originating in Canada, and as such, we were not able to use it for this work\footnote{https://developers.tiktok.com/doc/research-api-codebook/}.
\subsubsection{Instagram}

Until August 14, 2024, Meta provided access to its CrowdTangle platform\footnote{https://about.fb.com/news/2023/11/new-tools-to-support-independent-research/}, which we used to collect Instagram data from January 1, 2022, to August 10, 2024. During this period, data collection followed a daily scraping schedule. After CrowdTangle was discontinued on August 14, 2024, we transitioned to a custom scraping solution, similar to the approach described by Abrahams~\cite{alexei_social_2023}, operating on a weekly basis.


\subsubsection{Facebook}

From early 2022 until mid-August 2024, Facebook data was collected via Meta’s CrowdTangle API. Following its discontinuation, unlike Instagram where we deployed a custom scraper, developing a scraper for Facebook proved more technically challenging and time-consuming. As a result, Facebook data collection was paused after August 14, 2024. We plan to resume collection and backfill the missing period (August 2024 to January 2025) once a feasible solution is implemented.

Meta has since introduced the Meta Content Library\footnote{\url{https://transparency.meta.com/researchtools/meta-content-library}}, a restricted-access tool that operates within a secure environment, incompatible with external pipelines. This limitation highlights the growing challenges of maintaining continuous, cross-platform data access as platform policies evolve.

\subsubsection{BlueSky}
For Bluesky, we utilize the official API\footnote{\url{https://docs.bsky.app/}} to collect public posts in a structured and efficient manner. The platform’s open protocol and clear documentation allow for streamlined integration into our crawler system. As a result, Bluesky data collection has been stable, fast, and reliable, requiring minimal post-processing. This makes it one of the most technically straightforward sources in our multi-platform pipeline. As our newest platform, we currently only collect influencer and politician data.

\subsubsection{YouTube}
We used the YouTube Data API\footnote{\url{https://developers.google.com/youtube/v3}} to collect metadata associated with videos posted by accounts in our seed list. This included post-level details such as titles, descriptions, publication timestamps, and engagement metrics like views and likes. To ensure temporal continuity and capture engagement activity close to the time of posting, we implemented a frequent and regular collection schedule. \\

Each record for each platform is annotated with a timestamp to indicate when the data was collected, independently of platform-provided metadata. This timestamp enables downstream analyses to account for time lags between post publication and data collection, supporting normalization of engagement metrics and cross-platform comparisons.

We also note that our dataset does not include content from other platforms such as Reddit, Mastodon, nor content from fringe platforms such as Parler or Truthsocial. But survey data collected by the Media Ecosystem Observatory indicates that these platforms account for a vanishing fraction of the Canadian public. To a first approximation, all the  major conversations and communities in Canadian political discourse are represented on the major mainstream social media platforms covered in our dataset. 

\section{Indexing Pipeline}

Handling multi-platform social media data requires a robust storage and processing architecture capable of reconciling heterogeneous data structures. Our pipeline begins with platform-specific media crawlers that write raw JSON data to cloud storage (Amazon S3\footnote{\url{https://aws.amazon.com/pm/serv-s3/}}). Keeping the raw JSON files is a critical design choice in our architecture. Rather than selecting specific fields for direct insertion into a database, we store full raw responses to preserve all metadata returned by the platforms. This allows for future re-indexing, schema updates, or downstream analyses that may require fields not initially considered. It also provides long-term flexibility: while the current pipeline uses Elasticsearch\footnote{\url{https://www.elastic.co/}} for indexing and querying, storing raw JSON enables easy migration to other storage or processing systems without data loss or format conversion overhead. 

A scheduled Apache Airflow\footnote{\url{https://airflow.apache.org/}} workflow orchestrates both the execution of platform-specific crawlers and the downstream transfer of raw data into Elasticsearch, a distributed search and analytics engine optimized for indexing large-scale semi-structured data. Each platform is assigned its own Elasticsearch index, preserving original schema elements for detailed, platform-specific querying.

To enable unified cross-platform analysis, a parallel normalization process is applied during ingestion. This process involves a custom Python-based transformation layer that maps heterogeneous schema fields to a common schema. For instance, while TikTok encapsulates all textual metadata in a single desc field, YouTube separates content into title and description, and Twitter maps text to the message field. These are all consolidated into a unified field. This allows researchers to perform full-text analysis across platforms. In addition, we embed contextual seed metadata (e.g., political affiliation, geographic location, content category) directly into each record, enabling filtered analysis by entity-level attributes.

By maintaining both platform-specific and normalized indices, our infrastructure supports dual modes of analysis: detailed per-platform investigations, and integrated multi-platform comparisons. This design allows for both granularity and generality in social media research, accommodating diverse methodological approaches from network analysis to topic modeling and content tracking.

\subsubsection{Semantic Indexing and Vector Database Integration}
In addition to storing and indexing raw and normalized data, our architecture incorporates a vector database layer to enable semantic retrieval and advanced content analysis. For each social media post, a dense vector representation (embedding) is computed using transformer-based language models. These embeddings are stored alongside the original post metadata in a dedicated vector index.

This setup allows us to perform high-efficiency similarity searches, such as retrieving posts semantically related to a given query or identifying clusters of discourse around shared topics. It also supports downstream tasks such as stance detection, misinformation tracking, and cross-platform narrative diffusion.

To support more advanced exploration, we store the results of unsupervised analyses—such as topic modeling and clustering—in a vector database, alongside post-level semantic embeddings. This setup allows researchers to search not only by keywords, but also by semantic similarity, enabling the discovery of related content across platforms and time.

By maintaining both a traditional keyword-based index and a semantic layer, our system supports hybrid queries that combine lexical, structural, and semantic criteria. This dual indexing approach increases flexibility for both qualitative exploration and computational workflows, especially in use cases involving narrative tracking, similarity-based retrieval, or longitudinal theme analysis.

\subsection{Data Access}
To ensure structured and uniform access to processed data and analysis results, we developed a set of RESTful APIs using FastAPI\footnote{\url{https://fastapi.tiangolo.com/}}. These endpoints allow authenticated users to query, retrieve, and analyze data.

The Data API provides endpoints for accessing seed lists, aggregated content statistics, timeline data, full-text search results, and normalized data, offering a consistent interface for content and metadata retrieval.

The system also provides an Analysis API that supports advanced computational tasks, including semantic similarity search, temporal pattern matching, and network-level comparisons. Network analyses leverage both graph neural network (GNN)-based embeddings and classical network metrics such as PageRank, modularity, and community detection. This enables multi-dimensional exploration of actors, content, and interactions across platforms. By standardizing access to these enriched data layers, the API allows researchers to perform consistent and reproducible analyses across a range of research questions.

All APIs support JSON input and output formats, enabling seamless integration with external scripts, dashboards, or Jupyter-based workflows. This modular API-first design simplifies access for both technical and non-technical users and supports reproducibility by ensuring that analysis procedures are centralized and parameterizable.

\begin{figure}
    \centering
    \includegraphics[width=0.7\linewidth]{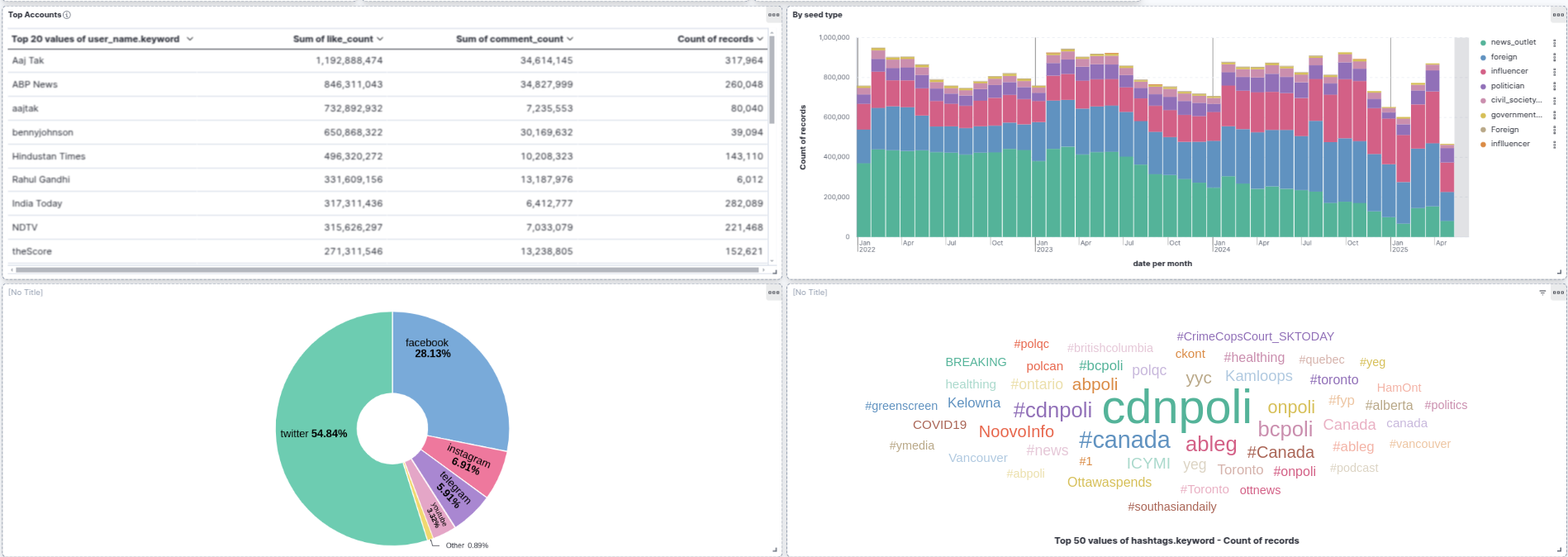}
    \caption{Kibana dashboard example}
    \label{fig:dashboard_example}
\end{figure}

Additionally, visual dashboards were created using Kibana\footnote{\url{https://www.elastic.co/kibana}} \ref{fig:dashboard_example}, which connects directly to the Elasticsearch back-end. These dashboards provide accessible interfaces for exploring trends, timelines, and content engagement patterns, without requiring programming expertise.

While Kibana provides powerful tools for internal data exploration and dashboarding, it assumes familiarity with the underlying data structure and query logic. To support broader accessibility, we developed a custom web interface called the MEO Insights Hub\footnote{\url{https://www.cdmrn.ca/data-and-resources}}. This frontend provides partner researchers with simplified access to the most commonly used analyses, including timelines and content filtering, and integrates selected Kibana dashboards where appropriate.

Together, this layered access architecture provides both high-level overview tools and low-level analytic flexibility, making the observatory a practical and extensible platform for multi-platform social media research.

\subsection{Monitoring Layer}
To ensure the reliability and stability of our data collection and processing pipeline, the observatory includes a dedicated monitoring layer. Each component in the system—including media crawlers, data normalization scripts, and indexing jobs—logs its status and performance metrics to Amazon CloudWatch\footnote{\url{https://aws.amazon.com/cloudwatch/}}. This centralized logging infrastructure allows for real-time tracking of failures, latencies, and throughput across all stages of the pipeline.

In addition to system-level monitoring, the observatory also maintains a crawler history tracker to manage and validate data completeness—particularly for platforms where data access is limited and backfilling is often required. For each crawler run, metadata is recorded to track which time intervals have been successfully collected. When gaps are detected—e.g., if a specific date range was missed or partially scraped—the system automatically flags these periods for historical recovery through back-scraping procedures.

Furthermore, Airflow orchestrates and logs all recurring pipeline jobs, ensuring observability over batch tasks such as scheduled exports, topic modeling runs, or vector updates. This multi-layered monitoring system provides operational transparency and allows for rapid response to pipeline disruptions, data inconsistencies, or platform-specific changes, ensuring minimal data loss and downtime.

\section{Exploratory Data Analysis}

The exploratory data analysis (EDA) presents a high-level overview of the dataset’s composition, including the distribution of records across platforms and entity types. This section aims to describe the dataset’s structure and coverage to inform subsequent analyses.

\begin{table}[ht]
\centering
\caption{Total and average (per seed) number of posts by platform and actor type. Averages are shown in parentheses.}
\small 
\renewcommand\cellalign{lc}
\renewcommand\cellgape{\Gape[4pt]}
\begin{tabular}{@{}lcccccc@{}}
\toprule
\textbf{Platform} & \textbf{News} & \textbf{Foreign} & \textbf{Influencer} & \textbf{Politician} & \textbf{CSO} & \textbf{Gov.} \\
\midrule
X/Twitter     & \makecell{7 322 094\\(10 961)} & \makecell{2 440 052\\(15 062)} & \makecell{6 207 101\\(5 957)} & \makecell{1 281 554\\(866)} & \makecell{720 391\\(1 149)} & \makecell{289 404\\(1 507)} \\
Facebook    & \makecell{4 820 888\\(6 352)}  & \makecell{3 455 607\\(47 337)} & \makecell{39 289\\(348)}     & \makecell{779 431\\(747)}   & \makecell{268 722\\(575)}   & \makecell{155\\(78)}         \\
Instagram   & \makecell{402 375\\(989)}      & \makecell{700 420\\(8 439)}    & \makecell{250 424\\(416)}    & \makecell{523 768\\(358)}   & \makecell{370 184\\(607)}   & \makecell{53 322\\(476)}     \\
YouTube     & \makecell{238 630\\(1 550)}    & \makecell{714 321\\(10 505)}   & \makecell{56 274\\(316)}     & \makecell{25 319\\(91)}     & \makecell{61 464\\(121)}    & \makecell{10 858\\(85)}      \\
TikTok      & \makecell{45 476\\(771)}       & \makecell{16 378\\(1 170)}     & \makecell{143 961\\(554)}    & \makecell{8 379\\(68)}      & \makecell{20 234\\(146)}    & \makecell{1 345\\(673)}      \\
Telegram    & \makecell{2 088\\(2 088)}      & \makecell{1 899 389\\(39 571)} & \makecell{72 794\\(3 309)}   & –                            & \makecell{5 536\\(1 845)}   & –                            \\
Bluesky     & –                              & –                               & \makecell{74 201\\(412)}     & \makecell{14 567\\(79)}     & –                            & –                            \\
\bottomrule
\end{tabular}

\label{tab:platform_distribution}
\end{table}

Table~\ref{tab:platform_distribution} reports both the total number of posts and the average number of posts per seed (i.e., per account) across platforms and actor types. While raw post volume highlights platforms like X/Twitter and Facebook as dominant in terms of overall activity, the average values offer a more nuanced view of engagement intensity. For instance, although TikTok and Telegram appear marginal in total volume, the average number of posts per actor is relatively high in some categories—particularly among influencers and foreign entities—suggesting high activity from a small number of highly engaged accounts. On the other hand, platforms like Instagram and YouTube display more moderate average activity, reflecting their slower content cycles.

By presenting both total and normalized figures, the table allows us to better understand platform dynamics and avoid misinterpreting size for influence. When using this dataset for comparative studies, particularly across platforms, such normalization can help account for representational imbalances.

\begin{figure}
    \centering
    \includegraphics[width=0.99\linewidth]{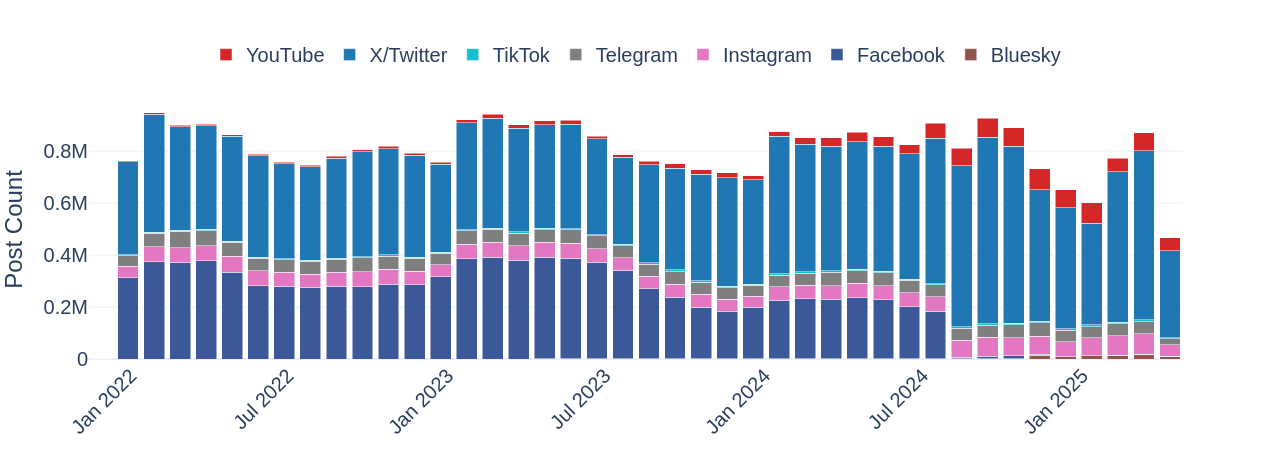}
    \caption{Timeline by platform}
    \label{fig:timeline}
\end{figure}

A time series chart of posts by platform can be found in Figure \ref{fig:timeline}, representing the data collected over two years. Variations in activity levels may correspond to platform-specific data availability, scraping limitations, or external factors such as significant events or campaigns, as addressed in Section \ref{section:discussions}.

\section{Applications}

Since the launch of Media Ecosystem Observatory, we have used the infrastructure not only to collect and store data, but also to carry out continuous, applied analyses of the Canadian information ecosystem. These analyses serve both as demonstrations of system capabilities and as contributions to ongoing public discourse about media, politics, and platform dynamics in Canada. In this section, we present selected examples that illustrate how the system has been used to monitor political communication, media coverage, and cross-platform dynamics.

\subsubsection{Meta News Ban}
In response to the Canadian government's Online News Act, Meta began blocking news content from its platforms for Canadian users on August 9, 2023. This policy change, referred to here as “Meta ban”, removed the visibility of all Canadian news content on Facebook and Instagram. Because our observatory infrastructure was already actively collecting Facebook data before the ban—and continued doing so afterward—we were uniquely positioned to monitor its effects in real time. This continuity enabled longitudinal analyses of the ban’s impact.

Our data shows that engagement with the Facebook Pages of Canadian news outlets dropped dramatically following the ban—approximately 64\% for national outlets and 85\% for local outlets. Nearly half of local news outlets stopped posting entirely within four months of the ban's implementation. By contrast, user engagement with political Facebook Groups remained largely stable, suggesting that politically engaged users continued to rely on the platform for discussion.

Interestingly, many users circumvented the ban by posting screenshots of Canadian news articles in political Groups. Although fewer in number than pre-ban news link shares, these screenshots generated similar levels of engagement, indicating continued demand for news content even under restrictive platform conditions. At the same time, the volume of misinformation links shared within political Groups declined, a counterintuitive result that may reflect changes in user behavior or moderation algorithms. This infrastructure enabled our team to produce timely, data-driven assessments of the ban's effects; see our preliminary report for more details \cite{parker_when_2024}.

\subsubsection{Federal Elections}

During the most recent Canadian federal elections, the infrastructure described in this paper played a central role in supporting both large-scale monitoring and qualitative analysis of online political discourse. The system enabled researchers to track and compare activity across multiple platforms in near real time, while also providing structured access to historical data for post-hoc verification and in-depth analysis.

This infrastructure underpinned a series of weekly situation reports, which offered timely insights into trends in political messaging, engagement, and issue salience. These publicly available reports can be accessed at: \url{https://www.cdmrn.ca/weekly-updates/}.

Using topic modeling methods such as BERTopic \cite{grootendorst2022bertopic}, we identify and track dominant themes in political communication and media coverage. These themes are visualized through time series plots, keyword drift analyses, and semantic clustering diagrams. Because our dataset spans multiple platforms—including X/Twitter, YouTube, TikTok, Instagram, and Bluesky—we are able to observe how topics evolve, diverge, or synchronize across distinct digital spaces.
\begin{figure}
    \centering
    \includegraphics[width=0.5\linewidth]{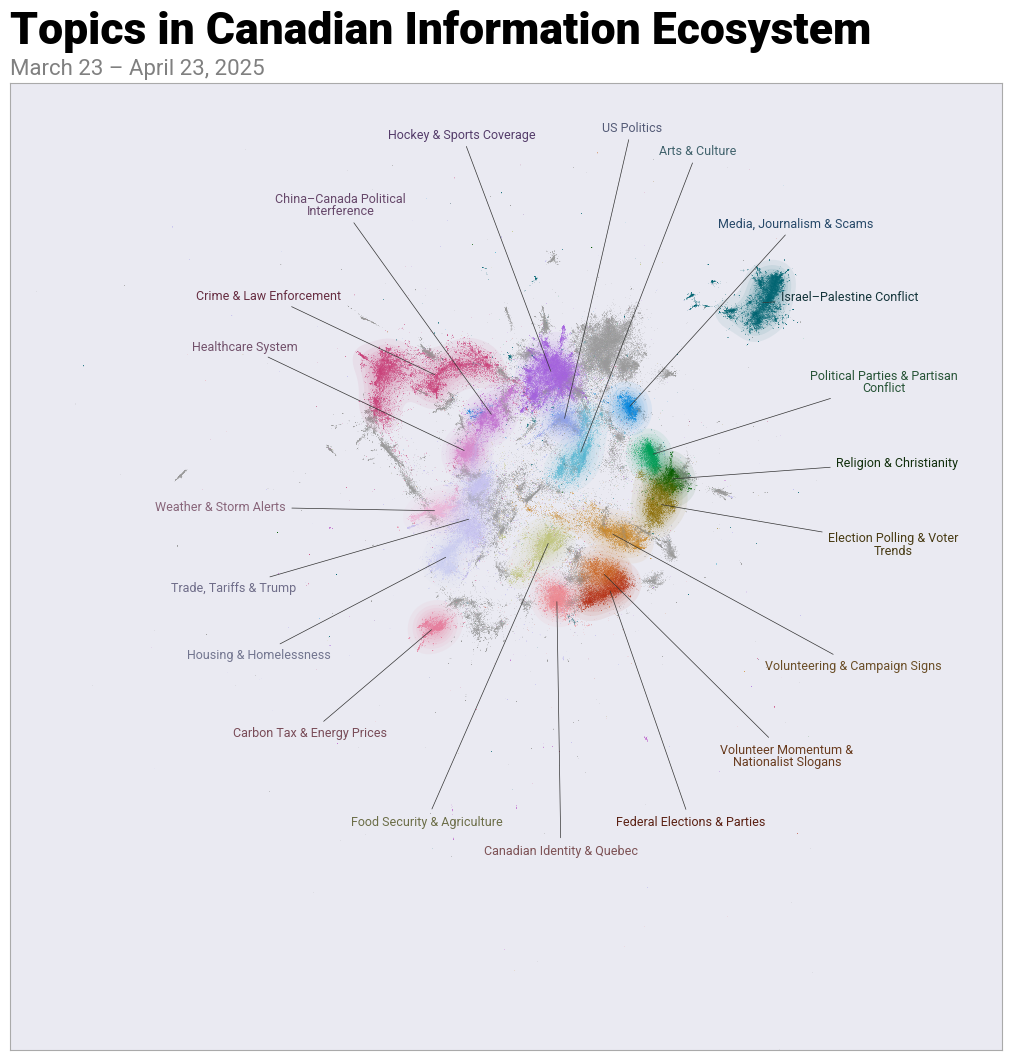}
    \caption{Topic analysis during federal elections 2025}
    \label{fig:topics}
\end{figure}





More recently, we’ve leveraged semantic embeddings and vector search to support exploratory research. Each post is embedded as a dense vector, enabling researchers to retrieve semantically similar content, detect narrative shifts over time, and examine coordinated messaging. These capabilities have been especially useful in comparing narratives during key political moments, when different parties or media outlets may engage in parallel but rhetorically distinct campaigns.

All of these analyses contribute to a growing series of public-facing outputs, including the Monthly Situation Reports, which offer a recurring snapshot of the health and structure of Canada’s online information ecosystem.

\section{Discussion and Conclusion}
\label{section:discussions}

\subsection{Limitations}
While the Canadian Media Ecosystem Observatory aims to provide a comprehensive, multi-platform dataset, several limitations remain due to platform restrictions, methodological constraints, and scope choices made during development.
Facebook coverage is the most prominent gap. Although we successfully collected Facebook data from early 2022 until Meta’s shutdown of the CrowdTangle API on August 14, 2024, we were not able to implement an alternative solution in time. Unlike Instagram, where we transitioned to a custom scraper, Facebook’s structure presented more complex technical challenges. We plan to fill the gap (August 2024–January 2025) in future iterations. Meta's new “Content Library” is not directly compatible with our pipeline due to its secure clean-room architecture.

Platform variability also affects data quality and consistency. For TikTok, Instagram (post CrowdTangle), and X/Twitter, data was collected via scraping techniques. These methods are less stable than official APIs and may be impacted by antiscraping mechanisms or platform-side changes. For example, Instagram data after August 2024 comes from a scraper, and while we have not observed systematic discrepancies, scraping introduces more variability and fragility compared to API-based access. X/Twitter scraping relies on public search functionality and may miss posts due to search index limitations and rate caps. Since our scrapers and seed list enrichment evolved over time, coverage windows may differ across platforms. To account for this, we include a crawled date field in all records, allowing researchers to assess collection timing and coverage.

Seed list constraints also introduce limitations. While we aimed for near-complete coverage of federal and provincial politicians, we excluded municipal politicians from towns under 100,000 residents for scalability. News outlet coverage is based on MediaCloud’s Canadian collections, supplemented by manual verification. Although extensive, some marginal outlets may still be missing.

Finally, replication challenges are inherent in social media research. Posts may be deleted, accounts made private, or engagement metrics may change over time. Although we retain metadata snapshots for each post, some temporal drift in metrics is inevitable. These limitations are common in digital trace research but should be considered when interpreting results based on our dataset.

\subsection{Ethical Considerations}

In response to the increasing difficulty of accessing platform data via official APIs, our infrastructure uses scraping selectively and responsibly to capture public content from verified or high-impact accounts. We explicitly avoid scraping private user data or personal communications.  Crawlers are rate-limited and metadata-only; media files are not collected. This approach is aligned with recent ethical guidelines \cite{brown2024webscrapingresearchlegal}, which offer a detailed framework for legal, ethical, and scientific best practices in web scraping for research purposes. Their work highlights the importance of transparency, data minimization, privacy risk assessment, and institutional review processes—all of which are considered in our design and data governance protocols. We advocate for ethically justified, transparency-oriented scraping practices in the face of platform-imposed opacity. Such approaches are critical to maintaining independent media research infrastructures in a post-API era.

\subsection{Future Directions}
While we are satisfied with the current state of the infrastructure, several directions remain for further development and refinement. Our current design prioritizes entity-based data collection, which results in a clean, well-structured dataset anchored to verified actors such as politicians and news outlets. This approach enables robust, comparative analysis across time and platforms.

To complement this, we have begun integrating event-based collection mechanisms, such as hashtag- and keyword-driven crawlers. These allow us to capture spontaneous and emerging phenomena—including viral campaigns, misinformation events, and protests—that fall outside the core entity seed list. To maintain the clarity of our main dataset, such data is ingested into separate, event-specific collections, as it tends to be noisier and requires more delicate analysis.

We have also implemented multilingual support, primarily through the normalization pipeline and NLP tools that account for both English and French content. In future versions, we aim to expand this to better support other languages.Additional enhancements in progress include automated content classification (e.g., stance, emotion detection) and automated entity detection to extend our seed list.

\subsection{Conclusion}
This paper presented the design and implementation of the Canadian Media Ecosystem Observatory, a modular, multi-platform system for monitoring and analyzing political and media discourse across social media. By combining large-scale data collection, cross-platform normalization, semantic indexing, and programmatic access via APIs and dashboards, the observatory offers a powerful resource for researchers and practitioners seeking to understand how information flows in Canada’s digital public sphere.

Yet the observatory is not a static product, but an evolving infrastructure. Maintaining it requires constant adaptation to platform changes, emerging narratives, and shifting research needs. It is a system that thrives on continuous engagement—technical, analytical, and collaborative. We see this project not as a finished endpoint, but as an open, ongoing journey—one that invites further contributions, critical reflection, and interdisciplinary use.

%
%
%
\bibliographystyle{splncs04}
\bibliography{meo}

\begin{thebibliography}{10}
\providecommand{\url}[1]{\texttt{#1}}
\providecommand{\urlprefix}{URL }
\providecommand{\doi}[1]{https://doi.org/#1}

\bibitem{alexei_social_2023}
Abrahams, A.S.: Social media observatory (2023), \url{https://nostarch.com/social-media-observatory}

\bibitem{Ai_Gupta_Oak_Hui_Liu_Hirschberg_2024}
Ai, L., Gupta, S., Oak, S., Hui, Z., Liu, Z., Hirschberg, J.: Tweetintent@crisis: A dataset revealing narratives of both sides in the russia-ukraine crisis. Proceedings of the International AAAI Conference on Web and Social Media  \textbf{18}(1),  1872--1887 (May 2024). \doi{10.1609/icwsm.v18i1.31432}, \url{https://ojs.aaai.org/index.php/ICWSM/article/view/31432}

\bibitem{bossetta_cross-platform_2023}
Bossetta, M., Schmøkel, R.: Cross-platform emotions and audience engagement in social media political campaigning: Comparing candidates’ facebook and instagram images in the 2020 {US} election  \textbf{40}(1),  48--68 (2023). \doi{10.1080/10584609.2022.2128949}, \url{https://doi.org/10.1080/10584609.2022.2128949}, publisher: Routledge \_eprint: https://doi.org/10.1080/10584609.2022.2128949

\bibitem{brown2024webscrapingresearchlegal}
Brown, M.A., Gruen, A., Maldoff, G., Messing, S., Sanderson, Z., Zimmer, M.: Web scraping for research: Legal, ethical, institutional, and scientific considerations (2024), \url{https://arxiv.org/abs/2410.23432}

\bibitem{dogdu_detecting_2024}
Dogdu, E., Choupani, R., Sürücü, S.: Detecting {Political} {Polarization} {Using} {Social} {Media} {Data}. In: Han, H., Baker, E. (eds.) Next {Generation} {Data} {Science}. pp. 46--59. Springer Nature Switzerland, Cham (2024). \doi{10.1007/978-3-031-61816-1_4}

\bibitem{grootendorst2022bertopic}
Grootendorst, M.: Bertopic: Neural topic modeling with a class-based tf-idf procedure. arXiv preprint arXiv:2203.05794  (2022)

\bibitem{hase_adapting_2023}
Hase, V., ~, Karin, B., , Scharkow, M.: Adapting to {Affordances} and {Audiences}? {A} {Cross}-{Platform}, {Multi}-{Modal} {Analysis} of the {Platformization} of {News} on {Facebook}, {Instagram}, {TikTok}, and {Twitter}. Digital Journalism  \textbf{11}(8),  1499--1520 (Sep 2023). \doi{10.1080/21670811.2022.2128389}, \url{https://doi.org/10.1080/21670811.2022.2128389}

\bibitem{matassi_agenda_2021}
Matassi, M., Boczkowski, P.: An {Agenda} for {Comparative} {Social} {Media} {Studies}: {The} {Value} of {Understanding} {Practices} {From} {Cross}-{National}, {Cross}-{Media}, and {Cross}-{Platform} {Perspectives}. International Journal of Communication  \textbf{15}(0), ~22 (Jan 2021), \url{https://ijoc.org/index.php/ijoc/article/view/15042}

\bibitem{muric_covid-19_2021}
Muric, G., Wu, Y., Ferrara, E.: {COVID}-19 {Vaccine} {Hesitancy} on {Social} {Media}: {Building} a {Public} {Twitter} {Data} {Set} of {Antivaccine} {Content}, {Vaccine} {Misinformation}, and {Conspiracies}. JMIR Public Health and Surveillance  \textbf{7}(11),  e30642 (Nov 2021). \doi{10.2196/30642}, \url{https://publichealth.jmir.org/2021/11/e30642}, company: JMIR Public Health and Surveillance Distributor: JMIR Public Health and Surveillance Institution: JMIR Public Health and Surveillance Label: JMIR Public Health and Surveillance Publisher: JMIR Publications Inc., Toronto, Canada

\bibitem{parker_when_2024}
Parker, S., Park, S., Pehlivan, Z., Abrahams, A., Desblancs, M., Owen, T., Phillips, J., Bridgman, A.: When journalism is turned off: Preliminary findings on the effects of meta’s news ban in canada (2024). \doi{10.31235/osf.io/eqn45}, \url{https://osf.io/eqn45}

\bibitem{pehlivan2021archiving}
Pehlivan, Z., Thi{\`e}vre, J., Drugeon, T.: Archiving social media: the case of twitter. In: The past web: Exploring web archives, pp. 43--56. Springer (2021)

\bibitem{perakakis_social_2019}
Perakakis, E., Mastorakis, G., Kopanakis, I.: Social {Media} {Monitoring}: {An} {Innovative} {Intelligent} {Approach}. Designs  \textbf{3}(2), ~24 (Jun 2019). \doi{10.3390/designs3020024}, \url{https://www.mdpi.com/2411-9660/3/2/24}

\bibitem{ronzhyn_defining_2023}
Ronzhyn, A., Cardenal, A.S., Batlle~Rubio, A.: Defining affordances in social media research: {A} literature review. New Media \& Society  \textbf{25}(11),  3165--3188 (Nov 2023). \doi{10.1177/14614448221135187}, \url{https://doi.org/10.1177/14614448221135187}

\bibitem{Shao_2016}
Shao, C., Ciampaglia, G.L., Flammini, A., Menczer, F.: Hoaxy: A platform for tracking online misinformation. In: Proceedings of the 25th International Conference Companion on World Wide Web - WWW ’16 Companion. p. 745–750. WWW ’16 Companion, ACM Press (2016). \doi{10.1145/2872518.2890098}, \url{http://dx.doi.org/10.1145/2872518.2890098}

\bibitem{steel2024invasion}
Steel, B., Parker, S., Ruths, D.: The invasion of ukraine viewed through large-scale analysis of tiktok  (2024)

\bibitem{sun_social_2024}
Sun, Y., Jia, R., Razzaq, A., Bao, Q.: Social network platforms and climate change in china: Evidence from {TikTok}  \textbf{200},  123197 (2024). \doi{10.1016/j.techfore.2023.123197}, \url{https://www.sciencedirect.com/science/article/pii/S004016252300882X}

\bibitem{wiedemann_concept_2023}
Wiedemann, G., Münch, F.V., Rau, J.P., Kessling, P., Schmidt, J.H.: Concept and challenges of a social media observatory as a {DIY} research infrastructure. Publizistik  \textbf{68}(2),  201--223 (Sep 2023). \doi{10.1007/s11616-023-00807-6}, \url{https://doi.org/10.1007/s11616-023-00807-6}

\bibitem{Yang2022Botometer101}
Yang, K.C., Ferrara, E., Menczer, F.: Botometer 101: Social bot practicum for computational social scientists. Journal of Computational Social Science  \textbf{5},  1511--1528 (2022). \doi{10.1007/s42001-022-00177-5}, \url{https://rdcu.be/cUvyT}

\bibitem{yang_regional_2024}
Yang, Z., Imouza, A., Touzel, M.P., Amadoro, C., Desrosiers-Brisebois, G., Pelrine, K., Levy, S., Godbout, J.F., Rabbany, R.: Regional and {Temporal} {Patterns} of {Partisan} {Polarization} during the {COVID}-19 {Pandemic} in the {United} {States} and {Canada} (Jul 2024). \doi{10.48550/arXiv.2407.02807}, \url{http://arxiv.org/abs/2407.02807}, arXiv:2407.02807 [cs]

\bibitem{yarchi_political_2021}
Yarchi, M., Baden, C., Kligler-Vilenchik, N.: Political polarization on the digital sphere: A cross-platform, over-time analysis of interactional, positional, and affective polarization on social media  \textbf{38}(1),  98--139 (2021). \doi{10.1080/10584609.2020.1785067}, \url{https://doi.org/10.1080/10584609.2020.1785067}, publisher: Routledge \_eprint: https://doi.org/10.1080/10584609.2020.1785067

\end{thebibliography}

\end{document}